\documentclass[aps, 10pt, prd,
              notitlepage, twocolumn, superscriptaddress,
              nofootinbib,floatfix]{revtex4-1}

\usepackage{aas_macros}
\usepackage{amsmath}
\usepackage{amssymb}
\usepackage{amsfonts}
\usepackage[utf8]{inputenc}
\usepackage[T1]{fontenc}
\usepackage{mathrsfs}
\usepackage{anyfontsize}

\usepackage[linktocpage,breaklinks]{hyperref}
\usepackage[usenames,dvipsnames]{xcolor}

\usepackage{tensor}
\usepackage{bm}

\usepackage{graphicx}
\usepackage{epsfig}
\usepackage{epstopdf}

\usepackage{natbib}
\usepackage{hyperref}

\hypersetup{colorlinks=true,
            citecolor=black,
            linkcolor=black,
            urlcolor=black}

\begin{document}

\title{Finite size effects on the light curves of slowly-rotating neutron stars}

\author{Hajime Sotani}
\email{sotani@yukawa.kyoto-u.ac.jp}
\affiliation{Division of Science,
National Astronomical Observatory of Japan, 2-21-1 Osawa, Mitaka,
Tokyo 181-8588, Japan}

\author{Hector O. Silva}
\email{hector.okadadasilva@montana.edu}
\affiliation{eXtreme Gravity Institute, Department of Physics,
Montana State University, Bozeman, Montana 59717 USA}

\author{George Pappas}
\email{gpappas@auth.gr}
\affiliation{Dipartimento di Fisica, ``Sapienza" Universit\'{a} di Roma \&
Sezione INFN Roma1, Piazzale Aldo Moro 5, 00185, Roma, Italy}
\affiliation{Department of Physics, Aristotle University of Thessaloniki,
Thessaloniki 54124, Greece}

\date{\today}

\begin{abstract}
An important question when studying the light curves produced by hot spots on
the surface of rotating neutron stars is, how might the shape of the light
curve be affected by relaxing some of the simplifying assumptions that one would
adopt on a first treatment of the problem, such as that of a point-like spot.
In this work we explore the dependence of light curves on the size and shape
of a single hot spot on the surface of slowly-rotating neutron stars.
More specifically, we consider two different shapes for the hot spots (circular
and annular) and examine the resulting light curves as
functions of the opening angle of the hot spot (for both) and width (for the latter).
We find that the point-like approximation can describe light curves reasonably well,
if the opening angle of the hot spot is less than $\sim 5^\circ$.
Furthermore, we find that light curves from annular spots are almost the same as
those of the full circular spot if the opening angle is less than $\sim
35^\circ$ independently of the hot spot's width.
\end{abstract}

%

\maketitle

\section{Introduction}
\label{sec:I}

Supernova explosions, the last act of massive stars, are the birth places of
neutron stars. During these processes, the density inside the star significantly
exceeds the standard nuclear density and the gravitational and magnetic fields
inside/around the star become much stronger than those in the Solar System
\cite{Shapiro:1983du}. Thus, the resulting neutron stars are the best natural
candidates for probing the physics under these extreme environments. In fact,
due to the saturation property of nuclear matter, it is challenging to
obtain information about matter in such high-density regions in Earth
laboratories, which leads to many uncertainties in the equation of state (EOS)
of neutron stars~\cite{Lattimer:2015nhk}. This difficulty may be overcome by observing neutron stars
themselves and/or the phenomena associated to them.
One of the main observational constraints on the EOS at supra-nuclear densities
comes from the discoveries of $\sim2M_\odot$ neutron
stars~\cite{Demorest:2010bx,Antoniadis:2013pzd,Cromartie:2019kug}.  Due to the
existence of such massive neutron stars one can exclude the soft EOSs, that have
expected maximum masses that are less than the observed neutron star masses.
Another constraint on the EOS is the information about the tidal deformability
obtained from the observation of the gravitational waves emitted by GW170817,
which is the binary neutron star
merger~\cite{TheLIGOScientific:2017qsa,Abbott:2018exr}.
From the obtained tidal deformability, the radius of a $1.4M_\odot$ neutron
star is constrained to be around $\approx 11$ km~\cite{Kumar:2019xgp} with an
maximum value of $13.6$ km~\cite{Annala:2017llu}, which suggests that some of
the relatively stiffer EOSs in the low density region should be ruled out.
These examples demonstrate that the observations of neutron stars and
astrophysical phenomena involving them
are essential for constraining the EOS of neutron star matter at high
densities~\cite{Ozel:2016oaf}.

Electromagnetic observations of astrophysical processes involving neutron stars
offer another avenue to probe the properties of these extreme objects.
Due to their strong gravitational field, radiation from the immediate
vicinity of neutron stars experiences gravitational light bending, the
magnitude of which is controlled by the star's mass and radius, allowing (in principle)
these parameters to be inferred and consequently be used to shed light on the underlying
EOS~\cite{Watts:2016uzu,Watts:2018iom,Watts:2019lbs,Weih:2019rzo}.
Prime systems for such studies include accreting millisecond pulsars and
x-ray bursters. In both scenarios, parts of the star's surface are heated,
producing in this way hot spots (relative to the rest of the star).
These hot spots co-rotate with the star producing an observable x-ray flux which
is modulated by the star's spin frequency, called a light curve (or pulse
profile).
These light curves, encode information about the physical properties of the
hot spots such as e.g. its size, geometry, temperature distribution and spectra.
Moreover, they encode information of the spacetime curvature
around the star and thereby of the bulk properties of the star.
One of the main goals of the ongoing Neutron star Interior Composition ExploreR
(NICER)~\cite{2012SPIE.8443E..13G,2014SPIE.9144E..20A,2017NatAs...1..895G}
mission, is to measure light curves from a number of sources with unprecedented
time-resolution, constraining their masses and radii within $5-10\%$
accuracy in optimal cases~\cite{Lo:2013ava,Lo:2018hes,Miller:2014mca}
(see also e.g.~\cite{Psaltis:2013fha,Miller:2016kae,Bogdanov:2015tua}).

Several theoretical studies of the properties of
light curves from hot spots on the surface of rotating neutron stars have been
carried out in the past (see
e.g.~\cite{1983ApJ...274..846P,1995MNRAS.277.1177L,Miller:1997if,Beloborodov:2002mr,Poutanen:2008pg}).
The contribution of the various ingredients that affect the shape of a light
curve due to the rotation of the neutron star have been studied in several
instances.
For example, there have been calculations of the light curve that are taking
into account the Doppler factor and the time delay in the Schwarzschild
spacetime~\cite{Miller:1997if,Poutanen:2003yd,Poutanen:2006hw,Sotani:2018oad}
\footnote{The slow-rotating approximation without the Doppler factor and the
time delay is valid at least with $\sim 0.1$ Hz \cite{Sotani:2018oad}, while one can
see the significant deviation with a few hundred Hz \cite{Miller:1997if,Poutanen:2006hw}.}.
In other cases, light curves were calculated either within the Hartle-Thorne
approximation~\cite{Psaltis:2013zja} or in the numerically determined
spacetime of rapidly-rotating neutron stars~\cite{Cadeau:2004gm,Cadeau:2006dc}.
These works showed that the inclusion of effects such as
Doppler, aberration and gravitational time-delay in the Schwarzschild spacetime
can estimate relatively well the light
curve produced by moderately rapidly-rotating neutron stars with spin frequencies
$\lesssim 300$ Hz, above which the inclusion of stellar oblateness
is required~\cite{Cadeau:2004gm,Cadeau:2006dc,Morsink:2007tv,Nattila:2017hdb}.
In addition, since light curves from neutron stars also depend on the
spacetime geometry and on the gravitational theory, one can use light curve
observations to perform strong-field tests of general relativity
(GR),
although the effect of the gravitational theory may be degenerate
with uncertainties on the EOS~\cite{Sotani:2017bho,Sotani:2017rrt,Silva:2018yxz,Silva:2019leq}.

A simplifying assumption used in some of these studies is that of a point-like
    approximation for the hot spot, where its size is assumed to be negligible.
This approximation may be valid under some conditions, but in realistic
astrophysical scenarios one should
take into account the finite-size effects of the hot spot on the light curve
(e.g., ~\cite{Baubock:2015ixa,Lockhart:2019nch}).
Furthermore, hydromagnetic numerical simulations of accreting millisecond pulsars
reveal that these hot spots might not necessarily even be circular
and can instead have ring or crescent-moon-like
shapes~\cite{Kulkarni:2005cs,Kulkarni:2013sza} with implications on the
resulting light curve of known systems~\cite{Ibragimov:2009js,Kajava:2011fh}.

In this work, we examine in details the effects of the shape and size
of single hot spots, in the controlled scenario of slowly-rotating neutron
stars, which allows to single-out their effects in the light curve.
This paper is organized as follows. In Sec.~\ref{sec:II} we summarize
the theory behind light curve modelling of slowly-rotating neutron stars,
with special emphasis on how to deal with finite-size effects of the hot spot.
In Sec.~\ref{sec:III} we discuss how these light curves are calculated
numerically and present a simple approach to study ring-shaped hot spots.
Then, we investigate in detail the resulting light curves,
considering different hot-spot-observer arrangements and hot spot sizes. We
also contrast the differences between circular and annular hot spots and in
the latter case examine the validity of the point-like hot spot approximation.
Finally, in Sec.~\ref{sec:V} we summarize our findings and discuss possible
extensions of our work.
We use geometric units, $c=G=1$, where $c$ and $G$ denote the speed of light in
vacuum and the gravitational constant respectively, and use metric signature
$(-,+,+,+)$.

\section{Radiation flux from slowly-rotating neutron stars}
\label{sec:II}

For generality, we consider the following line element for a static, spherically symmetric spacetime
given by
\begin{equation}
  ds^2 = -A(r)\,dt^2 + B(r)\,dr^2 + C(r)\,(d\theta^2 + \sin^2\theta\, d\psi^2),
  \label{eq:line_element}
\end{equation}
where we focus on asymptotically flat spacetimes for which $A(r)\to 1$, $B(r)\to
1$, and $C(r)\to r^2$ as $r\to\infty$. In these coordinates, the circumference
radius, $r_c$, is given by $r_c^2\equiv C(r)$ at each radial position $r$.
This line element is general enough to describe neutron stars not only in
GR, but also in modified theories of gravity~\cite{Sotani:2017bho,Silva:2018yxz}. While in Sec.~\ref{sec:III}
the particular case of the Schwarzschild spacetime will be taken, in the present
section we develop a general approach for dealing with finite sized hot spots
that can readily be used outside of GR.

\begin{figure}
\begin{center}
\includegraphics[scale=0.4]{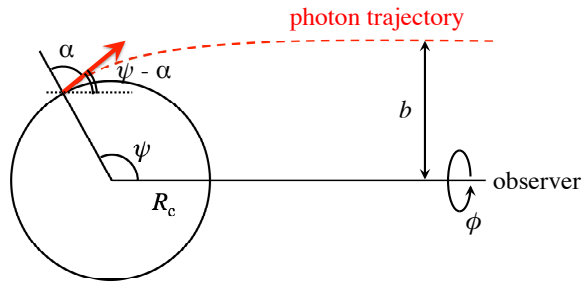}
\end{center}
\caption{
Geometry used to describe the emission of photons from a neutron star.
The dashed line indicates the trajectory of a photon, emitted with an
angle $\alpha$ (measured with respect to the normal to the star's surface)
from the star's surface $R_{c}$ [$R_{c} = C^{1/2}(R)$] from a point
located at $\psi$ (measured with respect to the line of sight to the observer).
The star's gravitational field bends the trajectory by an amount $\psi - \alpha$
and the photon is observed arriving with an impact parameter $b$.
The angle $\phi$ denotes the azimuthal angle in the observer's sky.
}
\label{fig:trajectory}
\end{figure}

Let us consider the trajectory of photons on this spacetime, initially emitted
from the star's surface located at $r=R$ and propagating outwards to an
observer located at a distance $D$ ($\gg R$) from the star.
The equation of motion for null geodesics can be obtained from the Euler-Lagrange
equation, i.e.,
\begin{equation}
  \frac{\partial{\cal L}}{\partial x^\mu} -\frac{d}{d\lambda}\left(\frac{\partial{\cal L}}{\partial \dot{x}^\mu}\right) =0,  \label{eq:EL}
\end{equation}
where ${\cal L}$ is the Lagrangian and the dot denotes a derivative with
respect to an affine parameter $\lambda$.
Using Eq. (\ref{eq:line_element}), we obtain
\begin{equation}
  2{\cal L} = -A\dot{t}^2 + B\dot{r}^2 + C\dot{\psi}^2, \label{eq:L}
\end{equation}
where, due to spherical symmetry, we can choose the photon trajectory to be on
the plane $\theta=\pi/2$ without loss of generality.
Then, combining Eqs. (\ref{eq:EL}) and (\ref{eq:L}), the angle at the stellar
surface $\psi(R)$ is given by~\cite{1983ApJ...274..846P}
\begin{equation}
    \psi(R) = \int_R^\infty \frac{dr}{C}\left[\frac{1}{AB}\left(\frac{1}{b^2} -
    \frac{A}{C}\right)\right]^{-1/2}, \label{eq:psiR}
\end{equation}
where
\begin{equation}
    \sin\alpha = b\,\sqrt{\frac{A(R)}{C(R)}}, \label{eq:b}
\end{equation}
with $b\equiv \ell/e$ being the impact parameter, defined in terms
of the photon's energy $e$ and angular momentum $\ell$, while $\alpha$ is the emission
angle as shown in Fig.~\ref{fig:trajectory}, measured relative to the normal
to the star's surface.

Using Eqs. (\ref{eq:psiR}) and (\ref{eq:b}), we can numerically calculate the
relation between $\psi(R)$ and $\alpha$ for a given radius $R$ (once $A$, $B$
and $C$ have been specified), where the value of $\psi$ increases as $\alpha$
increases.
This allows us to define a critical value of $\psi$
\begin{equation}
\psi_{\rm cri} \equiv \psi(\alpha = \pi/2)
\end{equation}
for a photon to reach the observer, where $\alpha=\pi/2$ means that the photon is emitted
tangentially with respect to the stellar surface. This in turn allows us to
introduce the notion of an \textit{invisible zone}, defined by $\alpha>\pi/2$
(or $\psi>\psi_{\rm cri}$) wherein photons cannot reach the observer.
We remark that the $\psi_{\rm cri}$ (and thus the invisible zone) also depends on the underlying theory of gravity.

Since light bending is a GR effect, $\psi_{\rm cri}$ increases as the
gravitational field becomes stronger. Thus, the value of $\psi_{\rm cri}$
increases (or conversely, the invisible zone decreases) as the compactness
${\cal C} \equiv M / R$ of the radiating object increases.
In the Schwarzschild spacetime, the invisible zone vanishes if
${\cal C} \geq 0.284$, implying that the whole surface of the star
is visible to the observer.
In the absence of an invisible zone, photon trajectories can whirl around the
star multiple times~\cite{1983ApJ...274..846P,Sotani:2018fhj} resulting in
multiple images of the star's surface.
These strong lensing effects do not happen for canonical neutron stars for which
$M/R \lesssim 0.2$ (see Fig.~3 in Ref.~\cite{Sotani:2018fhj}) which we will consider
hereafter.

How do we calculate the flux measured by an observer, 
that is coming from a single hot spot on the star's surface?
The area of the hot spot, $dS$, and the corresponding solid
angle on the observer's sky, $d\Omega$, are given by
\begin{align}
    dS &= C(R)\sin\psi \,d\psi \,d\phi,   \label{eq:dS}   \\
    d\Omega &= \frac{b\,db\,d\phi}{D^2},   \label{eq:dOmega}
\end{align}
where, recall, $D$ is the distance between the star and the observer, while
$\phi$ is an azimuthal angle in the observer's sky as shown in
Fig.~\ref{fig:trajectory}
The observed spectral flux, $dF_E$, inside $d\Omega$ is expressed as
\begin{equation}
 dF_E = I_E\,d\Omega,    \label{eq:dF}
\end{equation}
where $I_E$ is the specific intensity of the radiation measured in terms of the
energy $E$ measured at infinity by the observer.
It is however more convenient to write $I_E$ in terms of the specific intensity
$I_0(E_0,\alpha)$ measured by an observer at the vicinity of the stellar
surface. These two specific intensities are related as
\begin{equation}
I_E = A(R)^{3/2}I_0(E_0,\alpha),
\label{eq:i_transform}
\end{equation}
where $E_0$ is the photon energy measured by an observer at the vicinity of the
star's surface.

We can now combine Eq.~\eqref{eq:i_transform} with Eqs.~(\ref{eq:b}), (\ref{eq:dOmega}),
and~(\ref{eq:dF}) to obtain
\begin{equation}
  dF_E = \frac{A(R)^{1/2}C(R)}{D^2}I_0(E_0,\alpha)\sin\alpha\cos\alpha\frac{d\alpha}{d\psi}d\psi d\phi,
\end{equation}
which, upon integration over all energies $E$, leads to the observed bolometric
flux $dF$:
\begin{align}
    dF &= \int_0^\infty dF_E \, dE,
    \nonumber \\
       &= \frac{A(R)C(R)}{D^2}I_0(\alpha)\sin\alpha\cos\alpha\frac{d\alpha}{d\psi}d\psi d\phi.  \label{eq:dF1}
\end{align}
Here, $I_0(\alpha)\equiv \int I_0(E_0,\alpha)dE_0$ is the bolometric intensity
measured in the vicinity of the stellar surface. To derive Eq.
(\ref{eq:dF1}) we also converted  the emitted to observed energies
using the usual redshift relation
\begin{equation}
E = A(R)^{1/2} \, E_0.
\end{equation}
Although $I_0$ generally depends on the emission angle due to scattering as
photons propagate through the neutron star's atmosphere
~\cite{Zavlin:1996wd,2006ApJ...644.1090H,2012ApJ...749...52H,Salmi:2019pod}, we
consider for simplicity that the emission is uniform, i.e. $I_0=$ const., as in
the previous studies
(e.g.~\cite{Beloborodov:2002mr,Sotani:2017bho,Sotani:2018oad}).
The bolometric flux $F$ and the observed area of the hot spot $S$ are then given
by
%
\begin{align}
    F &= F_1 \iint_S \sin\alpha\cos\alpha\frac{d\alpha}{d\psi}d\psi\,d\phi\,,
    \nonumber \\
    S &= C(R) \iint_S \sin\psi \,d\psi \,d\phi\,,  \label{eq:F}
\end{align}
with
\begin{equation}
    F_1 \equiv \frac{I_0 A(R)C(R)}{D^2},
\end{equation}
and where $\iint_S$ means that the integration should be done only over the region
occupied by the hot spot.

\begin{figure}
\begin{center}
\includegraphics[width=0.6\columnwidth]{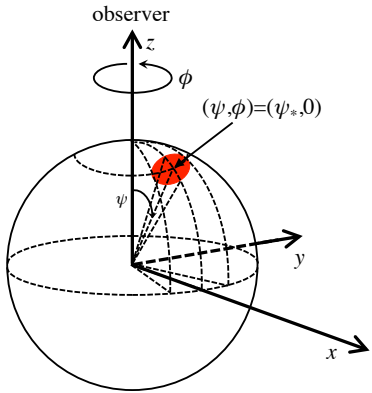}
\end{center}
\caption{
At any time, one can choose the coordinate where the center of hot spot is
located at $(\psi,\phi)=(\psi_*,0)$.
}
\label{fig:hotspot}
\end{figure}

Let us consider the case where the hot spot is circular, with size determined
by an opening angle $\Delta\psi$. As shown in Fig.~\ref{fig:hotspot}, we can
choose the coordinate at any time, where the center of the hot spot is located at
$(\psi,\phi)=(\psi_*,0)$. In these coordinates, the unit vector pointing to the
center of the hot spot, ${\bm n}$, and the unit vector pointing to an arbitrary
position of $(\psi,\phi)$, ${\bm x}$, are given by
${\bm n}=(\sin\psi_*,0,\cos\psi_*)$ and
${\bm x}=(\sin\psi\cos\phi,\sin\psi\sin\phi,\cos\psi)$.
The condition that the position of $(\psi,\phi)$ is inside the circular
hot spot is $\cos(\Delta\psi)\le{\bm n}\cdot{\bm x}$, which yields
\begin{equation}
  \cos(\Delta\psi) \le \sin\psi_*\sin\psi\cos\phi + \cos\psi_*\cos\psi.  \label{eq:HP}
\end{equation}

\begin{figure}[t]
\begin{center}
\includegraphics[width=0.65\columnwidth]{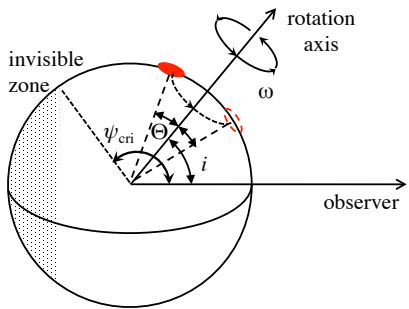}
\end{center}
\caption{
Illustration of the hot spot on a rotating star with angular velocity
$\omega$. The angle between the direction to the observer and the rotation
axis is $i$, while the angle between the center of the hot spot and the
rotation axis is $\Theta$. The shaded region denotes the invisible zone
whose boundary is determined by the angle $\psi_{\rm cri}$.
}
\label{fig:pulsar}
\end{figure}

The hot spot on a rotating neutron star with an angular velocity $\omega$
measured by the observer, is characterized by two angles, $i$ and $\Theta$, as
shown in Fig.~\ref{fig:pulsar}.
That is, $\Theta$ is the angle between the center of the hot spot and the
rotation axis, while $i$ is the angle between the direction to the observer
and the rotation axis.
Choosing $t=0$ when the hot spot is closest to the observer, the angular
position of the center of the hot spot, $\psi_*$, is given by
\begin{equation}
  \cos\psi_* = \sin i\sin\Theta \cos(\omega t) + \cos i\cos\Theta.  \label{eq:cos_psi}
\end{equation}

\begin{figure}[t]
\begin{center}
\includegraphics[width=0.75\columnwidth]{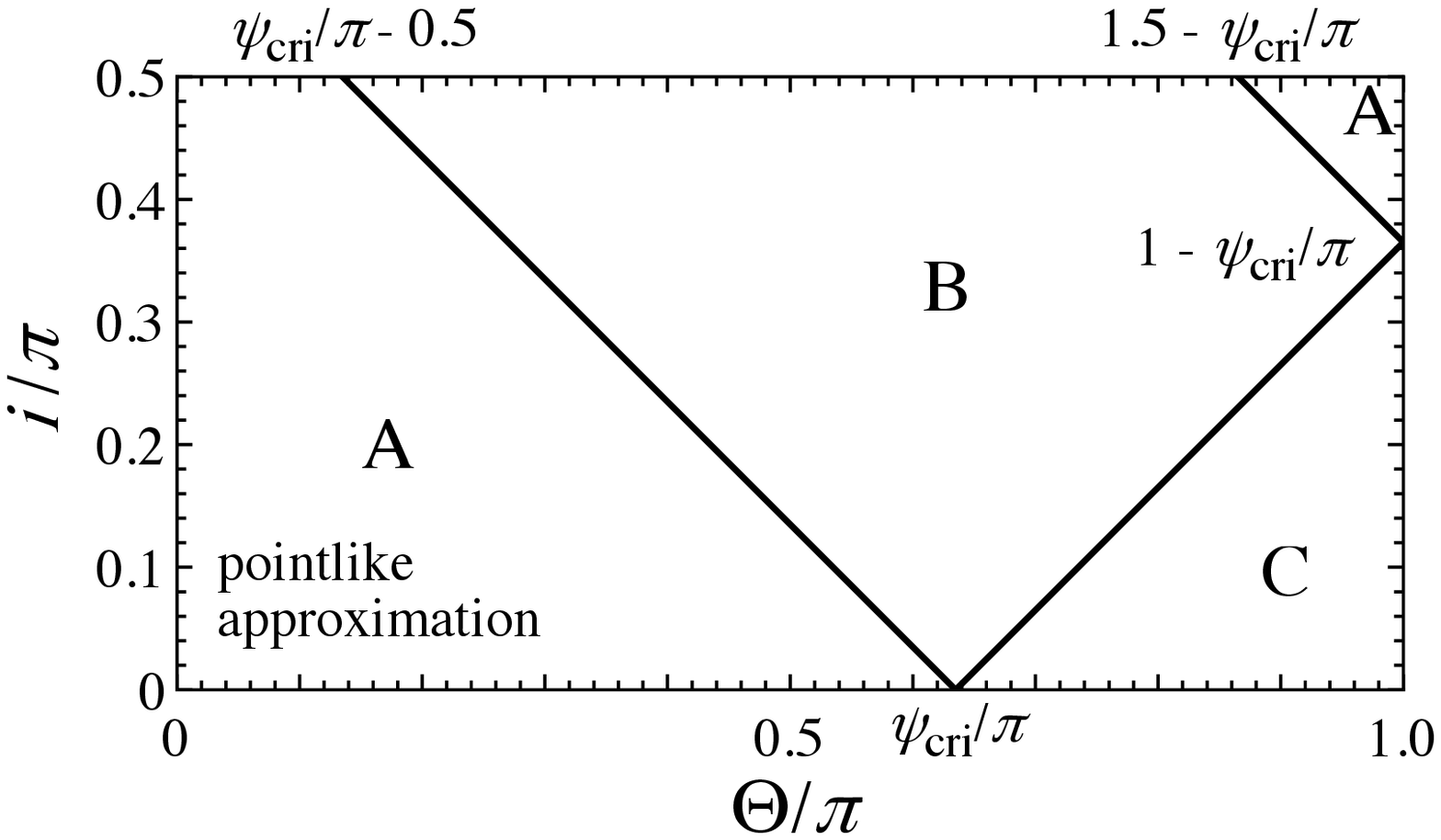}
\\
\includegraphics[width=0.75\columnwidth]{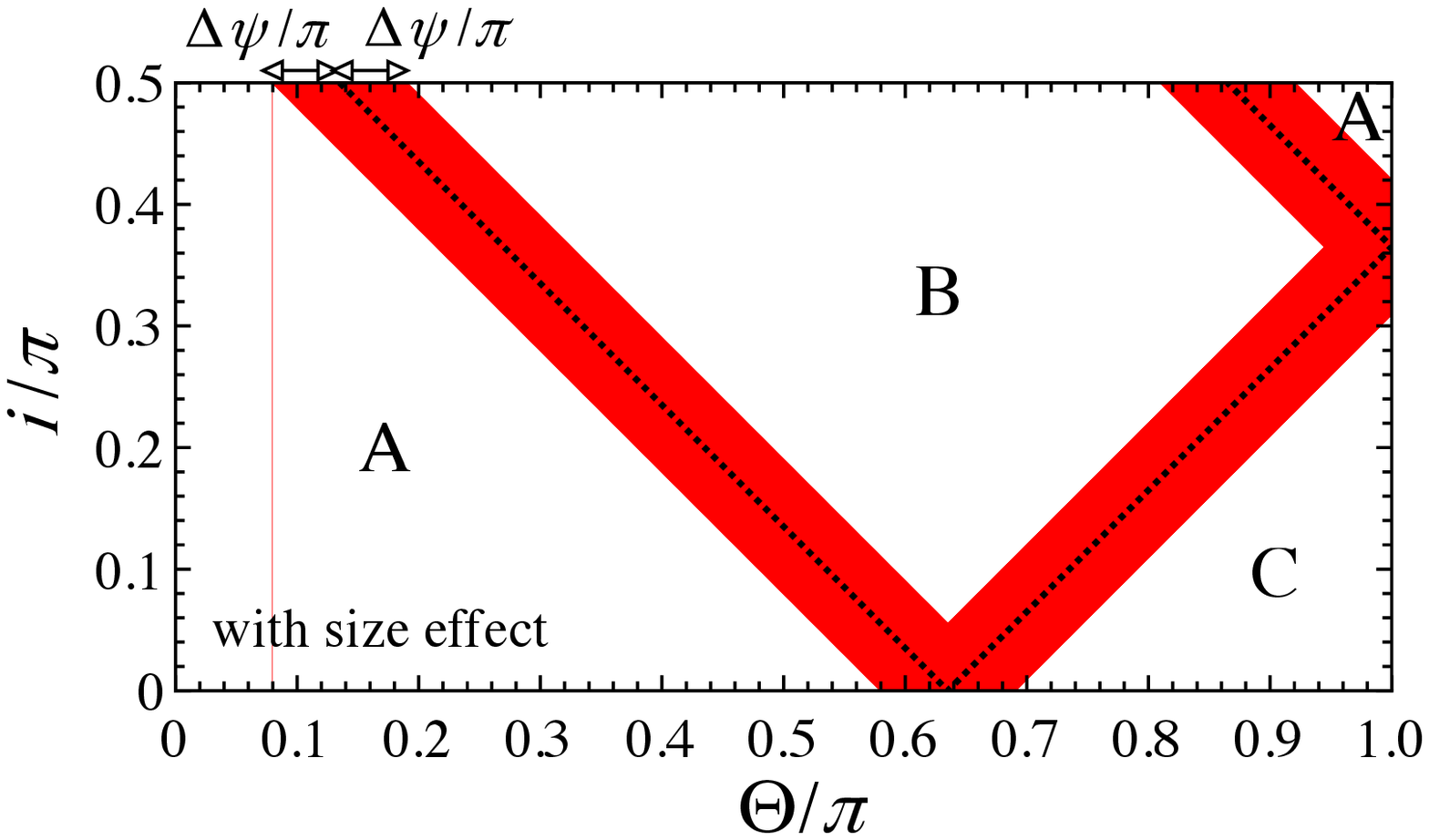}
\caption{
Visibility classification of the single hot spot, as a function of the angles of
$i$ and $\Theta$. The top panel corresponds to the case with a point-like
approximation, where three situations exist, i.e., (A) the hot spot can be always
observed, (B) the hot spot can enter the invisible zone for a fraction of the
period, and (C) the hot spot can not be observed in any time. The bottom panel
corresponds to the case with the effect of hot spot size, where the boundary of
the classification becomes blurred. This example is the case for
$\Delta\psi=10^\circ$. The colored region denotes the situation that a part of
the hot spot can enter the invisible zone.
}
\label{fig:class}
\end{center}
\end{figure}

Within the point-like approximation ($\Delta\psi \approx 0$)
(see the top panel of Fig.~\ref{fig:class}) one can classify
when the hot spot would be visible as a function of the angles of $i$ and $\Theta$,
as a function of $\psi_{\rm cri}$ as follows~\cite{Poutanen:2006hw}:
\begin{itemize}
  \item region A: the hot spot is always observed,
  \item region B: the hot spot enters the invisible zone for a fraction of the period,
  \item region C: the hot spot is invisible at any time.
\end{itemize}
When we include a finite size to the hot spot the boundaries between these
regions become blurred, as illustrated in the bottom panel of
Fig.~\ref{fig:class}.  In this panel, the shaded region corresponds to the
situation that a part of the hot spot can enter the invisible zone.

\section{Light curves of finite-sized hot spots}
\label{sec:III}

Until this point, our discussion has been very general and applicable for a
wide family of spacetimes whose line element can be written in the
form~\eqref{eq:line_element}.
From now on, we will consider the particular case of the Schwarzschild spacetime
to describe the star's exterior spacetime (motivated by Birkhoff's theorem in
GR) for which
\begin{equation}
 A(r) = 1-\frac{2M}{r}, \ \ B(r) = \frac{1}{A(r)}, \ \ C(r) = r^2.
\end{equation}
Nonetheless, we emphasize that the calculations and methods present here can
be applied to other static, spherically symmetric spacetimes.

Furthermore, to focus our attention on how the spot's area affects the light curve, we will
consider for simplicity only slowly-rotating neutron stars, where one can neglect
the rotational effects, such as Doppler shifts, aberration and the time
delay~\cite{Poutanen:2003yd,Poutanen:2006hw,Sotani:2018oad}.
In this case, the light curve obtained for $(\Theta,i)=(a,b)$ is the same as
that obtained for $(\Theta,i)=(b,a)$, a symmetry that arises due to Eq.~(\ref{eq:cos_psi}).
%
%
In practice, we will analyze light curves only for $\Theta=i$ and
more specifically show representative results for the cases with $\Theta=i=30^\circ$,
$45^\circ$, $60^\circ$, and $90^\circ$.
An additional symmetry of the light curve, is that its amplitude at $t/T$ for
$0.5\le t/T \le 1$ is the same as that at $1-t/T$, where $T$ is the
rotational period defined by $T \equiv 2\pi/\omega$.
In light of these properties, we will limit our calculation of the light curves
in the rotation phase interval $0 \le t/T\le 0.5$.
%
For our stellar models, we will consider two stars with $R=5M$ and $4M$, for
which $\psi_{\rm cri}$ are respectively $\psi_{\rm cri}=0.7191\pi$
($129.4^\circ$) and $\psi_{\rm cri}=0.8476\pi$ ($152.6^\circ$).

\begin{figure}
\includegraphics[scale=0.27]{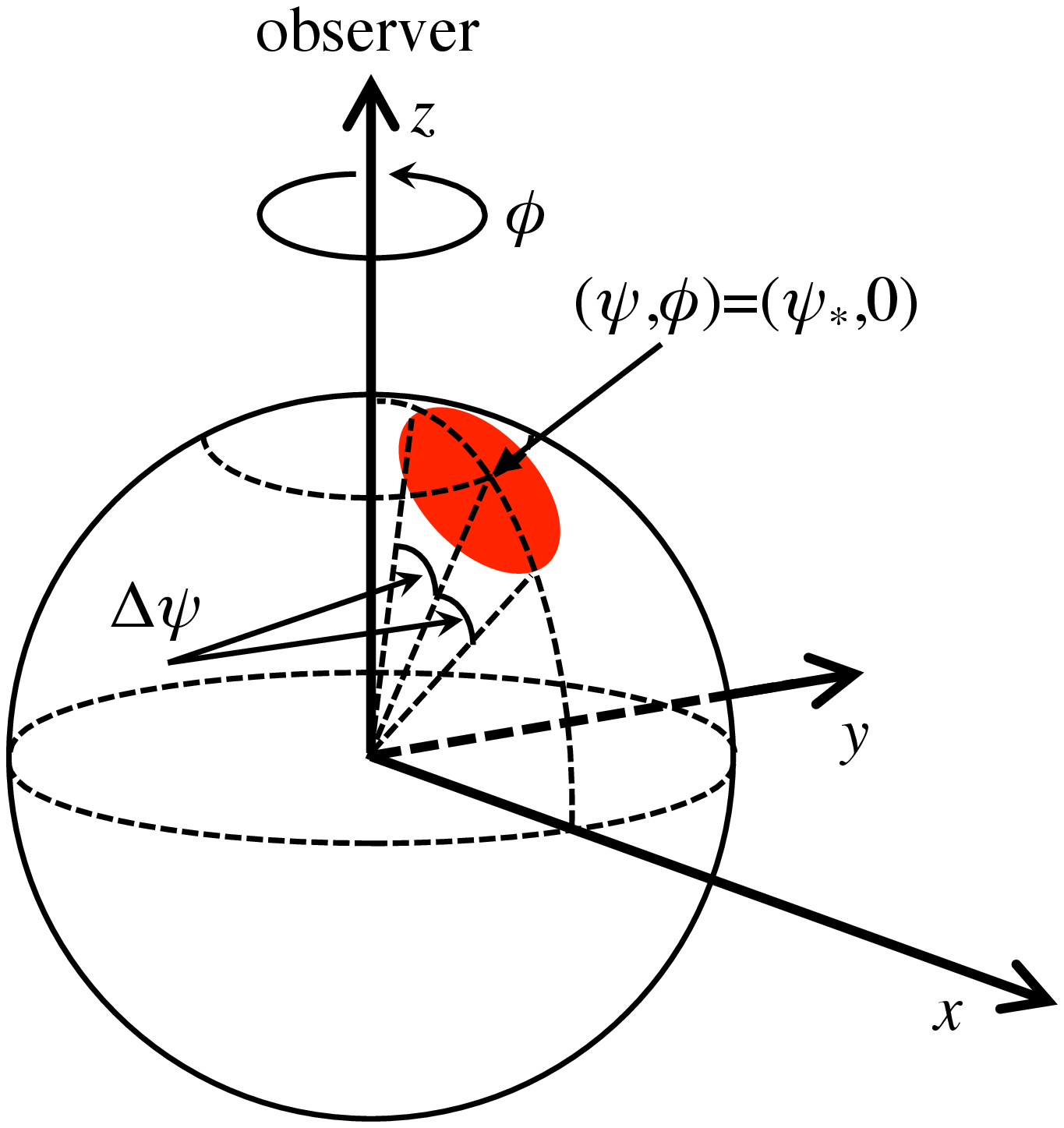}
\includegraphics[scale=0.27]{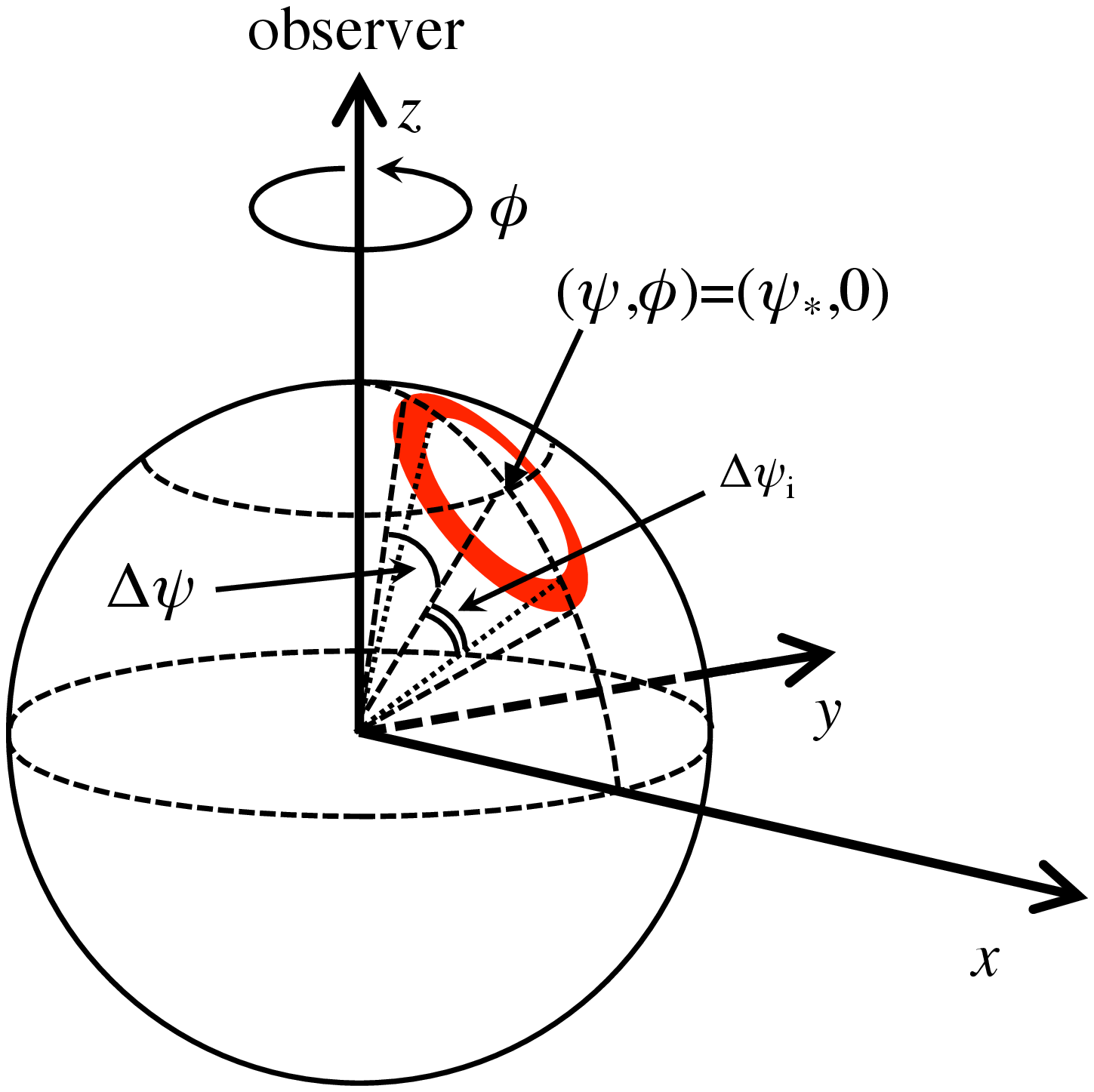}
\caption{
Illustration of two different geometries for the hot spot.
Left image: a circular hot spot with angular radius $\Delta \psi$.
Right image: an annular geometry with width $\Delta \psi - \Delta
\psi_i$.
In both cases, the center of the hot spot is chosen to be $\psi=\psi_*$.
}
\label{fig:HP}
\end{figure}

Considering these models, we will further assume two possible geometries for the hot spot.
First, in Sec.~\ref{sec:III-a}, we will consider circular hot spots with an
opening angle $\Delta\psi$, as shown in the left-hand-side of
Fig.~\ref{fig:HP}.
Next, in Sec.~\ref{sec:III-b}, we will study hot spots with an annular shape.
This geometry is obtained by removing from an otherwise circurlar hot spot (with
angular radius $\Delta \psi$), an internal circular region with opening angle $\Delta
\psi_i$. This situation is illustrated in the right-hand-side of
Fig.~\ref{fig:HP}.
In both cases, the hot spot's center is fixed at $\psi=\psi_*$.

To calculate the light curve of a circular hot spot we proceed as follows.
For any time instant, the center of the spot $\psi_*$, is determined from Eq.~(\ref{eq:cos_psi}).
Since the flux with a specific value of $\psi$ is independent of the value of $\phi$, the flux
from the hot spot is calculated from Eq. (\ref{eq:F}) as
\begin{equation}
    F(\Delta\psi)=2F_1 \int_{\psi_*-\Delta\psi}^{\psi_*+\Delta\psi} \phi(\psi)\sin\alpha\cos\alpha\frac{d\alpha}{d\psi}d\psi,
\end{equation}
where $\phi(\psi)$ is the boundary of the hot spot in the $\phi$ direction
determined from the constraint equation (\ref{eq:HP}) as a function of $\psi$.
As we explained previously, we can determine $\alpha(\psi)$ using Eqs.~\eqref{eq:psiR} and
\eqref{eq:b}, once a mass $M$ and radius $R$ has been specified for the neutron star.

In the case of an annular hot spot we exploit our assumptions of isotropic
emission and slow rotation. Under these assumptions, the flux coming from
the excised internal region (centered at $\psi_{\ast}$ with opening angle
$\Delta \psi_i$) can simply be \textit{subtracted} of the flux of the
total circular spot of opening angle $\Delta \psi$
, i.e.
\begin{align}
    F &= F(\Delta\psi) - F(\Delta\psi_i)
    \nonumber \\
    &= 2F_1 \int_{\psi_*-\Delta\psi}^{\psi_*+\Delta\psi} \phi(\psi)\sin\alpha\cos\alpha\frac{d\alpha}{d\psi}d\psi
    \nonumber \\
    &\quad - 2F_1 \int_{\psi_*-\Delta\psi_i}^{\psi_*+\Delta\psi_i} \phi_i(\psi)\sin\alpha\cos\alpha\frac{d\alpha}{d\psi}d\psi,
\end{align}
where $\phi_i(\psi)$ is the boundary of the inner circle.
%

\subsection{Circular hot spots}
\label{sec:III-a}

\begin{figure*}
\includegraphics[width=2\columnwidth]{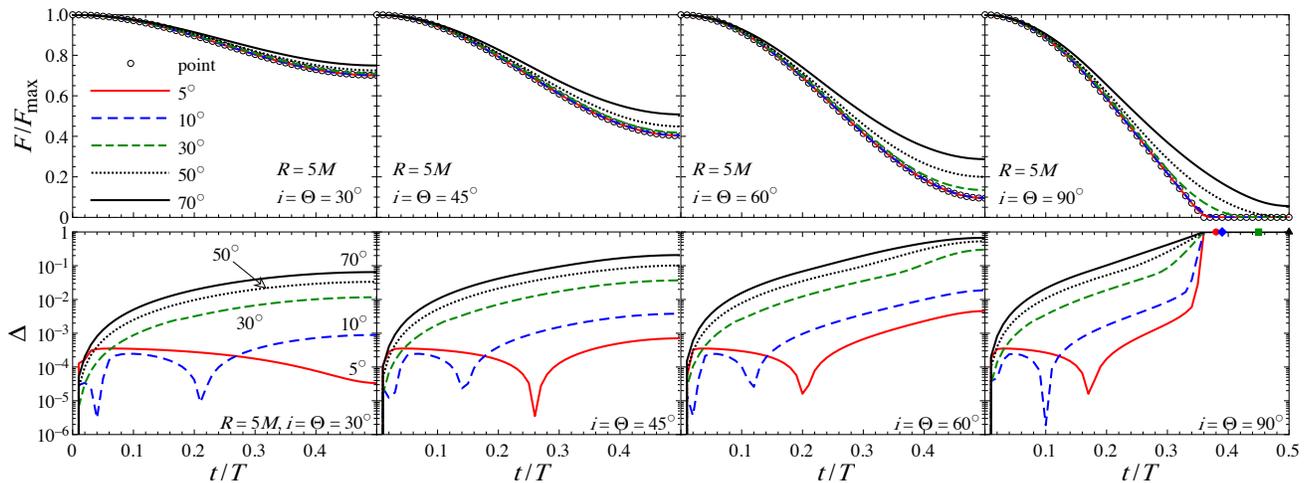}
\caption{
For the neutron star model with $R=5M$, the light curves calculated
considering the finite area of the hot spot are compared with that with the point-like spot
approximation.
In the top panels, the light curves with various opening angle
($\Delta \psi=5^\circ$, $10^\circ$, $30^\circ$, $50^\circ$, and $70^\circ$)
are shown with different lines, while the light curves with the point-like
approximation are shown with the open circles, where the all light curves
are normalized by their maximum values, i.e., the flux at $t/T=0$.
In the bottom panels, the relative deviations calculated by Eq. (\ref{eq:Delta})
are shown. The panels from left to right correspond to the case for
$i=\Theta=30^\circ$, $45^\circ$, $60^\circ$, and $90^\circ$.
For $i=\Theta=90^\circ$, the time when the observed flux becomes zero depends
on $\Delta \psi$, which is shown by the filled-circle, filled-diamond,
filled square, and filled-triangle for $\Delta\psi=5^\circ$, $10^\circ$, $30^\circ$,
and $50^\circ$ in the rightmost lower panel.
%
}
\label{fig:5Mi}
\end{figure*}

First, let us consider the light curves with the filled circle hot spots, as
shown in the left-hand side of Fig.~\ref{fig:HP}.
In the top panels of Fig.~\ref{fig:5Mi}, we show the light curves for the
neutron star model with $R=5M$, where the different lines correspond to the
results with different values of $\Delta\psi$ and for reference the results with
the point-like approximation are also shown by the open circles. The panels from
left to right correspond the results for the cases with $i=\Theta=30^\circ$,
$45^\circ$, $60^\circ$, and $90^\circ$.
%
To make explicit the role of the hot spot size on the light curve relative to
those calculated in the point-like approximation, we show the relative deviation
defined as
\begin{equation}
  \Delta \equiv \frac{|F/F_{\rm max} - F^{\rm (p)}/F_{\rm max}^{\rm (p)}|}{F/F_{\rm max}}, \label{eq:Delta}
\end{equation}
in the bottom panels of Fig.~\ref{fig:5Mi}. In Eq.~\eqref{eq:Delta}, $F/F_{\rm
max}$ and $F^{\rm (p)}/F_{\rm max}^{\rm (p)}$ denote respectively the observed
bolometric fluxes normalized by the maximum flux including the effect of spot's
finite size and that calculated using the point-like approximation.
From this figure, we observe (unsurprisingly) that the deviation from the result
calculated using the point-like approximation increases as the spot area
increases.
In particular, the deviation becomes significant when the hot spot
approaches the invisible zone.
This happens because, even if the center of the hot spot enters the invisible zone,
a part of it may still lay within the visible region.
Thus, \textit{when} exactly the flux is going to vanish completely depends strongly on the spot
size, as shown in the rightmost lower panel of Fig.~\ref{fig:5Mi}.
On the other hand, we also find that the light curves of hot spots with angular
radius up to $\Delta\psi\le 5^\circ$ are captured by the point-like
approximation within 1\% accuracy, if the center of the hot spot lays outside
the invisible zone.

\begin{figure}
\includegraphics[width=0.485\columnwidth]{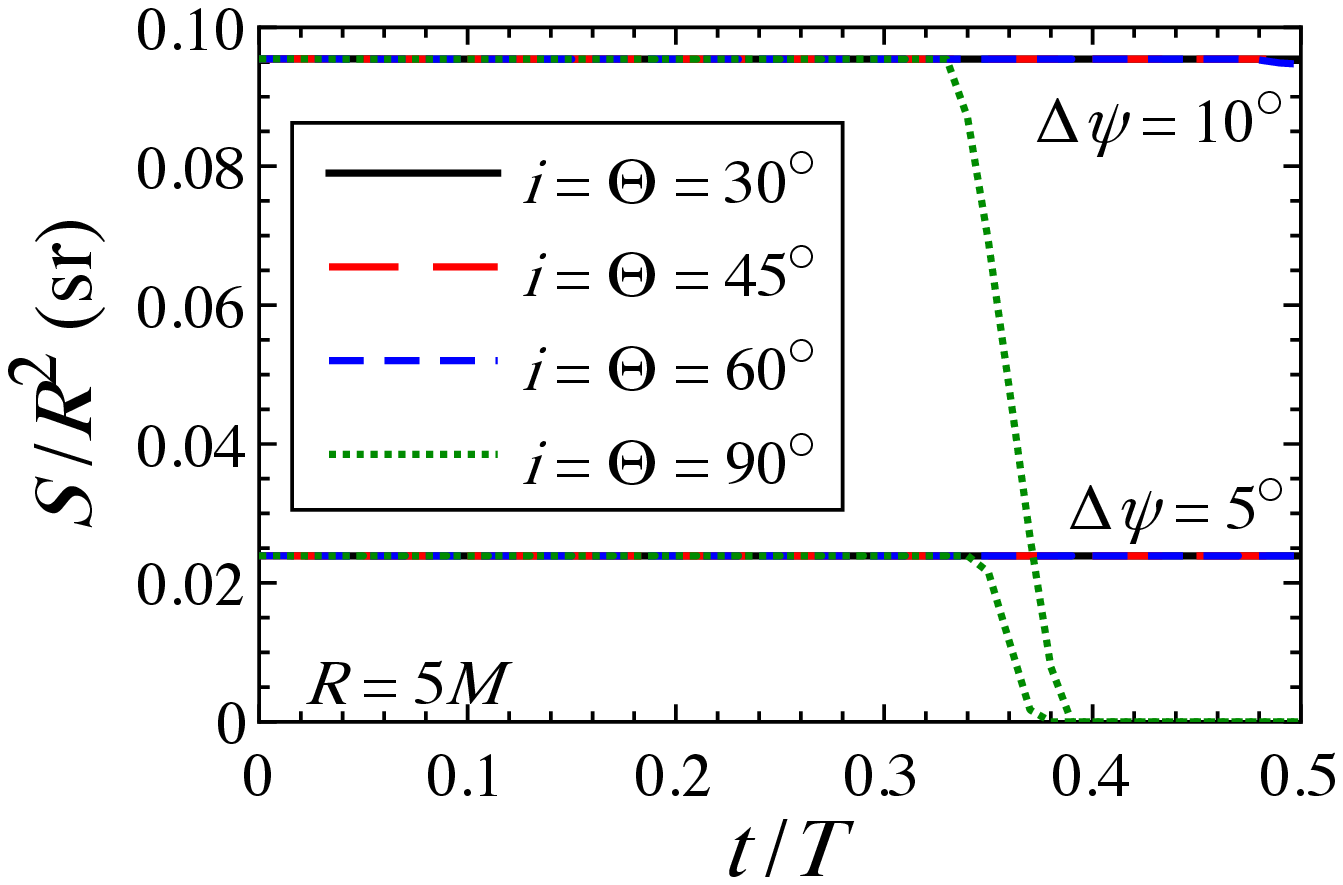}
\includegraphics[width=0.465\columnwidth]{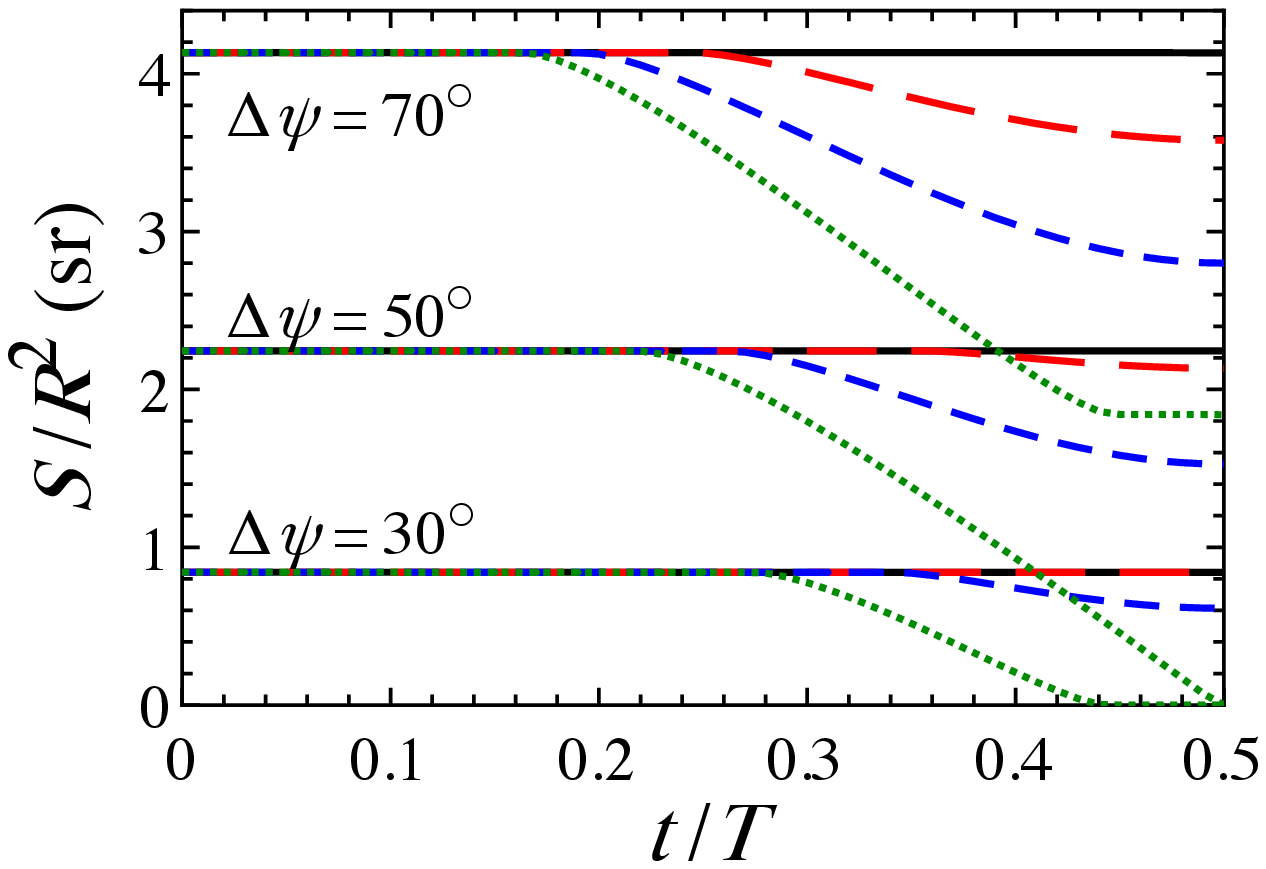}
\caption{
The visible area of the hot spot (normalized by $R^2$) for the neutron star
model with $R=5M$, for various combinations of $\Delta\psi$ and $i=\Theta$.
The significant decrease in the visible area indicates when a part of the hot spot
enters the invisible zone.
When $\Delta \psi = 70^{\circ}$, the visible hot area never drops to zero
regardless of the values of $i=\Theta$ considered.
%
}
\label{fig:5MS}
\end{figure}

In Fig.~\ref{fig:5MS} we show the spot area $S$ outside the invisible
zone, normalized by $R^2$.
We can see that when $i = \Theta$ and $\Delta \psi$ are large,
a portion of the hot spot's area enters the invisible zone earlier
over the course of the star's rotation.
We also observe that the spot area normalized by $R^2$ becomes
constant with $\Delta\psi=70^\circ$ and $i=\Theta=90^\circ$ for $t/T\sim
0.45-0.5$, where the invisible zone completely enters into the hot spot.
%
This happens due to the following. When the hot spot has a large opening angle,
$\Delta\psi$, it is more difficult for it to be completely eclipsed (i.e. to fall
completely in the invisible zone).
Notice that the opening angle of the invisible zone is given by
\begin{equation}
  \Delta\psi_{\rm cri} \equiv \pi - \psi_{\rm cri}, \label{eq:Dpsi}
\end{equation}
and it decreases as the compactness increases.
Consequently, if the opening angle of the spot $\Delta \psi$ is larger than
$\Delta\psi_{\rm cri}$, a fraction of the hot spot is \textit{always visible}.

\begin{figure}
\begin{center}
\includegraphics[width=0.75\columnwidth]{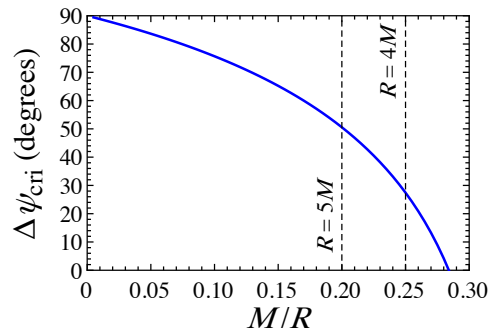}
\end{center}
\caption{
The critical value of the opening angle, $\Delta\psi_{\rm cri}$, given by
$\Delta\psi_{\rm cri}=\pi-\psi_{\rm cri}$ are shown as a function of $M/R$. For
the case when the opening angle $\Delta\psi$ is larger than $\Delta\psi_{\rm
cri}$ (the region above the line) the hot spot can not entirely enter into
the invisible zone. Two vertical dashed lines correspond to the values of $M/R$
for the stellar models with $R=5M$ and $4M$.}
%
\label{fig:MR-Dcri}
\end{figure}

In Fig.~\ref{fig:MR-Dcri}, we show the value of $\Delta\psi_{\rm cri}$ as a
function of $M/R$, where the two vertical dashed-lines correspond to the two 
stellar models with $R=5M$ (left line) and $4M$ (right line).
The values of $\Delta\psi_{\rm cri}$ are respectively $50.6^\circ$ and $27.4^\circ$
for stars with radii $R=5M$ and $R=4M$.
Hence, in Figs.~\ref{fig:5Mi} and~\ref{fig:5MS}, we can understand that the hot
spot with $\Delta\psi =70^\circ$ is always visible.
Moreover, we see that in principle, one could constrain the relation
between the stellar compactness and the opening angle of hot spot.
Specifically, if the flux vanishes (i.e. the hot spot
entirely enters into the invisible zone) one can constraint
the size of the opening angle of the hot spot to be
in the region below the line in Fig.~\ref{fig:MR-Dcri}.

\begin{figure*}
\begin{center}
\includegraphics[width=\textwidth]{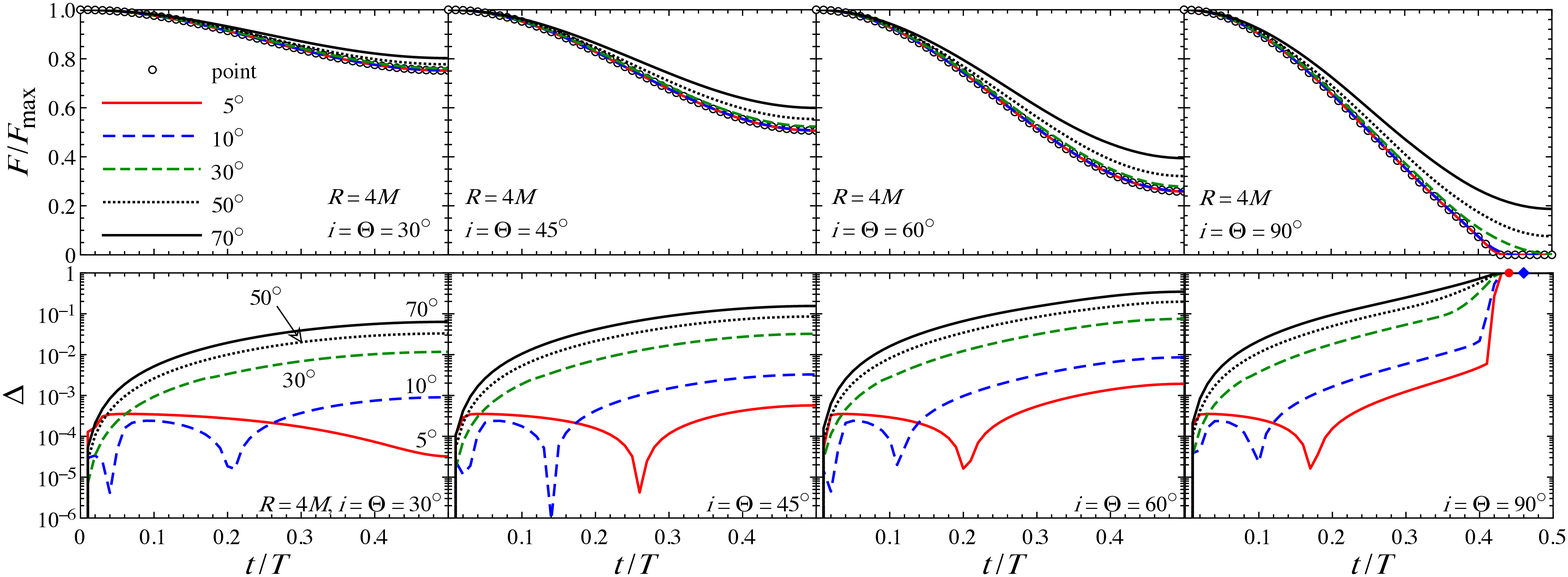}
\end{center}
\caption{
Same as Fig.~\ref{fig:5Mi}, but for the neutron star model with $R=4M$.
}
\label{fig:4Mi}
\end{figure*}

Similarly, we also calculated the light curves for the neutron star model
with $R=4M$, for which $\psi_{\rm cri}$ is larger in comparison to the previous
case ($R=5M$), i.e., the invisible zone becomes smaller.
In Fig.~\ref{fig:4Mi}, we show the light curves taking into account the finite
size of the hot spot together with the results using the point-like approximation
in the upper panels. The relative deviation [calculated using Eq.
(\ref{eq:Delta})] is shown in the lower panels.
The qualitative behavior of the light curve for $R=4M$ is very similar to that
for $R=5M$.
However, since the invisible zone is now smaller, the hot spot area hardly
enters it.
In fact, since $\Delta\psi_{\rm cri}=27.4^\circ$, even a relatively small
hot spot with $\Delta\psi=30^\circ$ is always visible when
$i=\Theta=90^\circ$.
Finally we see that, even for fairly compact neutron stars, the light curves
from a hot spot with $\Delta\psi\le 5^\circ$ can be well-captured using
point-like approximation, with relative errors of less than $1\%$, if the
center of the hot spot is outside the invisible zone.

\begin{figure}
\includegraphics[width=0.485\columnwidth]{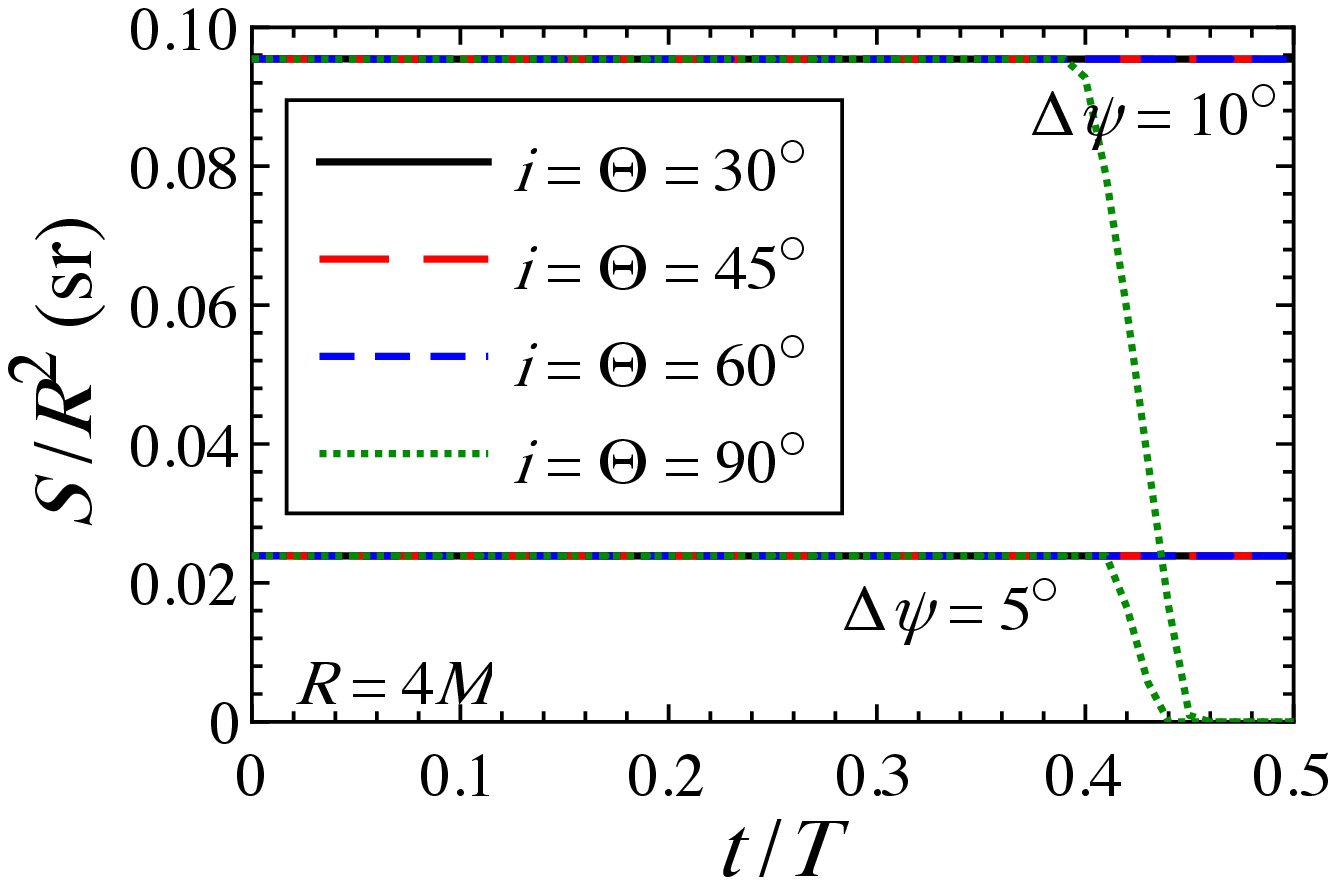}
\includegraphics[width=0.465\columnwidth]{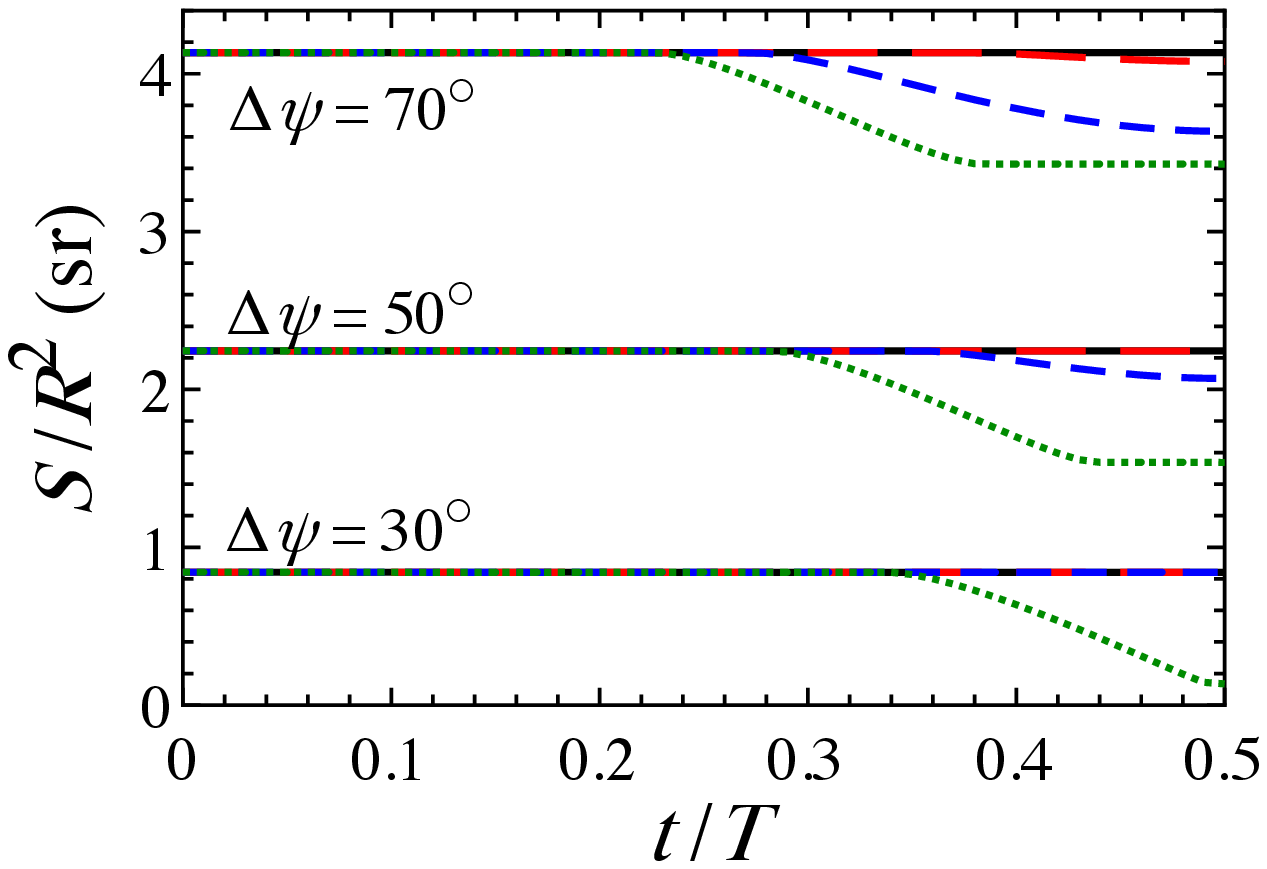}
\caption{
Same as Fig.~\ref{fig:5MS}, but for the neutron star model with $R=4M$.  This
more compact neutron star has a smaller invisible zone and therefore hot spot
sizes which would become invisible in the $R=5M$ case ($\Delta \psi =
30^{\circ}$ and $50^{\circ}$) are now always visible.
}
\label{fig:4MS}
\end{figure}

In Fig.~\ref{fig:4MS}, we show the visible hot spot area (again normalized by
$R^2$). We see that due to the smaller size of the invisible zone, it can
now \textit{completely lay within the hot spot} for $\Delta\psi=30^\circ$, $50^\circ$, and
$70^\circ$ when the hot spot is behind the neutron star. When this happens
the visible spot area becomes constant.

\subsection{Annular hot spots}
\label{sec:III-b}

Now let us consider a ring-shaped hot spot, as illustrated in the right-hand
side of Fig.~\ref{fig:HP}. In particular, we will consider hot spots with
$\Delta\psi=35^\circ$ and $70^\circ$, while varying the values of
$\Delta\psi_i$.

\begin{figure*}
\includegraphics[width=2\columnwidth]{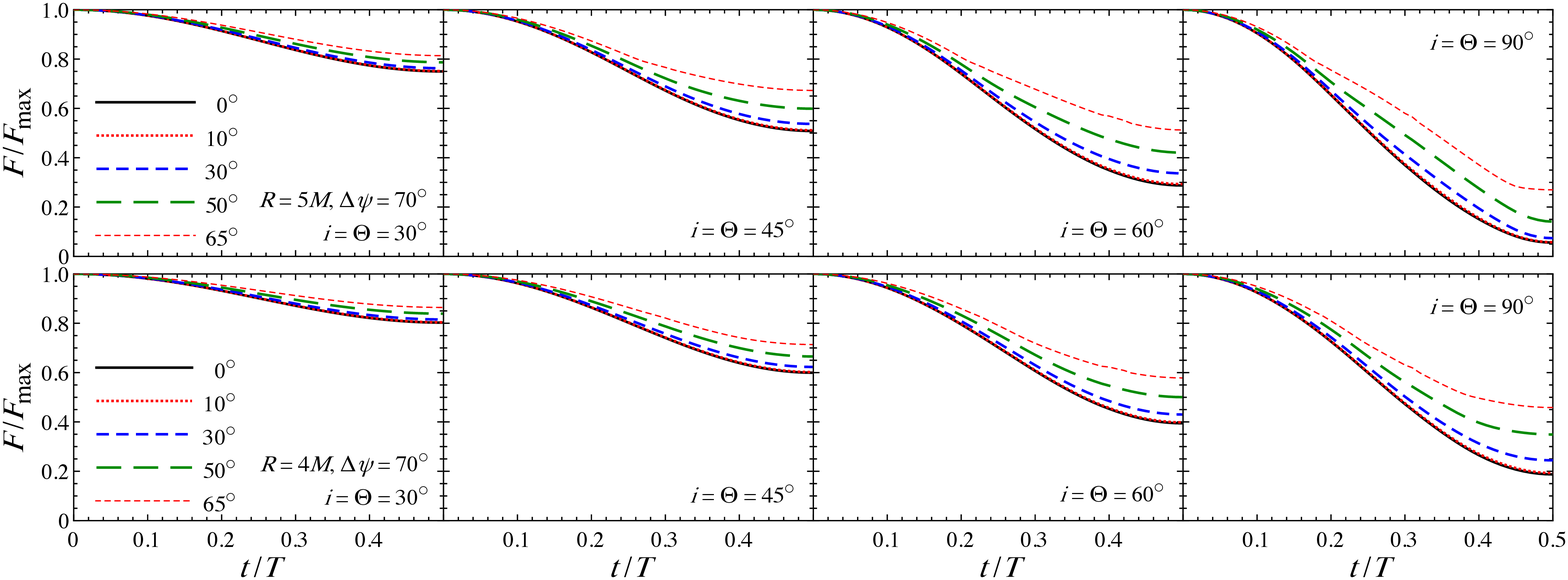}
\caption{
Light curves from the annular hot spot with $\Delta\psi=70^\circ$. The upper
and lower panels correspond to the observed flux normalized by the maximum flux
for the neutron star model with $R=5M$ and $4M$, respectively. The panels from
left to right correspond to the results for $i=\Theta=30^\circ$, $45^\circ$,
$60^\circ$, and $90^\circ$. In each panel, the different lines correspond to the
results with different values of $\Delta\psi_i$, i.e., $\Delta\psi_i=0^\circ$,
$10^\circ$, $30^\circ$, $50^\circ$, and $65^\circ$.
}
\label{fig:RD70}
\end{figure*}

In Fig.~\ref{fig:RD70},
we show the light curves from the spot with $\Delta\psi=70^\circ$ for the
neutron star with $R=5M$ in the upper panels and with $R=4M$ in the lower
panels, where the panels from left to right correspond to the results for
$i=\Theta=30^\circ$, $45^\circ$, $60^\circ$, and $90^\circ$. In each panel we
show the light curves for $\Delta\psi_i=0^\circ$, $10^\circ$, $30^\circ$,
$50^\circ$, and $65^\circ$, which correspond to $\Delta\psi_i/\Delta\psi = 0$,
$0.14$, $0.43$, $0.71$, and $0.93$ respectively. That is, the larger
$\Delta\psi_i/\Delta\psi$, the thinner is the radiating region of the hot spot.
The case of $\Delta\psi_i=0^\circ$ is shown only as a reference to facilitate
the comparison with the circular hot spot case discussed in the previous subsection.

\begin{figure}
\includegraphics[width=\columnwidth]{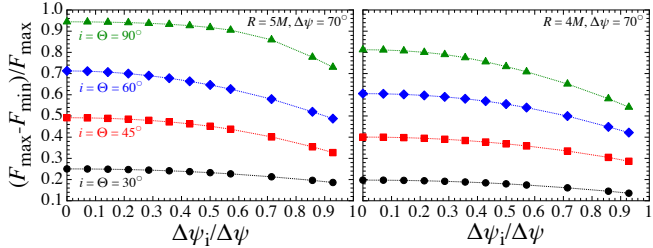}
\caption{
The ratio of the difference between the maximum and minimum flux to the maximum
flux is shown as a function of $\Delta\psi_i/\Delta\psi$ with
$\Delta\psi=70^\circ$, where the circles, squares, diamonds, and triangles
correspond to the results for $i=\Theta=30^\circ$, $45^\circ$, $60^\circ$, and
$90^\circ$. The left and right panels correspond to the results for the neutron
star model with $R=5M$ and $4M$.
}
\label{fig:Ratio70}
\end{figure}

In Fig.~\ref{fig:RD70}, we observe that the difference between the minimum flux,
$F_{\rm min}$, (at $t/T=0.5)$ and the maximum flux, $F_{\rm max}$, (at $t/T=0$)
decreases as $\Delta\psi_i$ increases in all the cases studied here.
 Additionally, we see that
such a difference becomes larger as the angle of $i=\Theta$ increases. To demonstrate
this behavior, we show the value of $(F_{\rm max}-F_{\rm min})/F_{\rm max}$ with
various angles of $i=\Theta$ in Fig.~\ref{fig:Ratio70}, where the left and right
panels correspond to the results for the neutron star model with $R=5M$ and
$4M$.
Furthermore, since it can happen that the invisible zone completely enters into
the internal region with the opening angle $\Delta \psi_i$ for the case of
$i=\Theta=90^\circ$, i.e., the hot spot is not eclipsed by the invisible zone when the hot spot
approaches the backside of the star, one can see in Fig.~\ref{fig:RD70} that the
shape of the light curve can deviate from that predicted by a filled hot
spot. This could be used (in principle) to distinguish between the
two different hot spot geometries.

\begin{figure*}
\includegraphics[width=2\columnwidth]{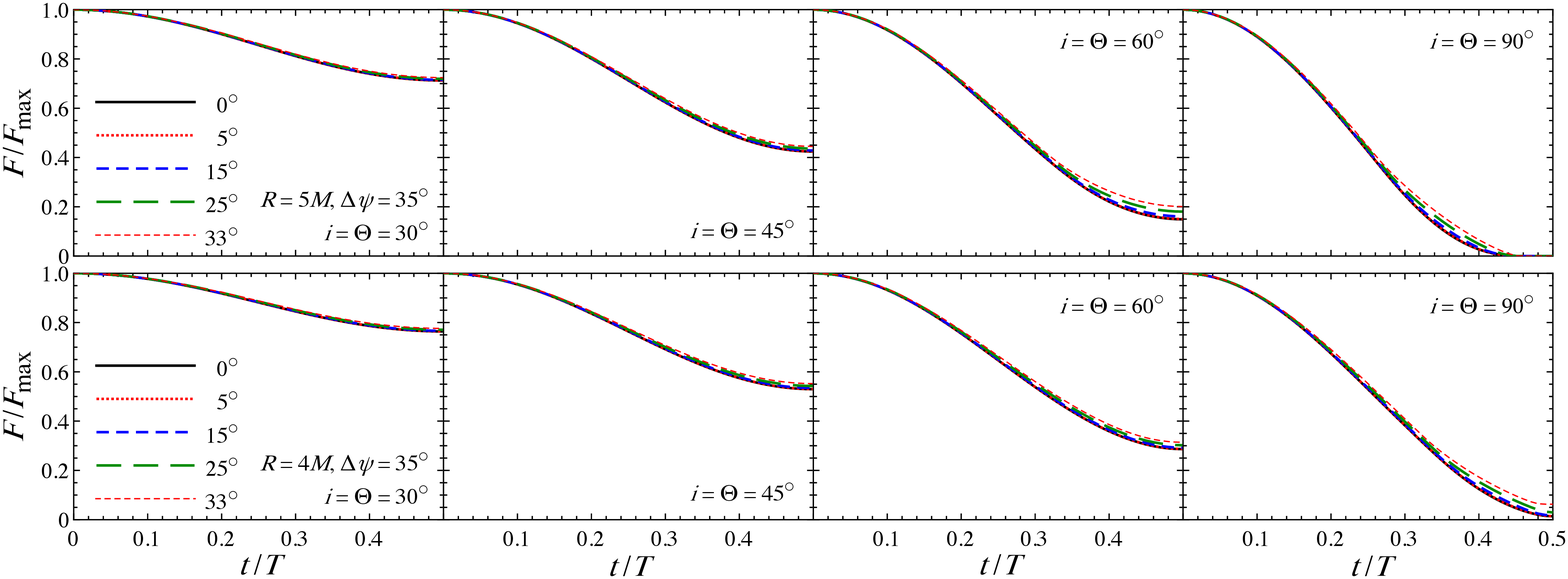}
\caption{
Same as Fig.~\ref{fig:RD70}, but for the neutron star model with $\Delta\psi=35^\circ$.
}
\label{fig:RD35}
\end{figure*}

\begin{figure}
\begin{center}
\includegraphics[width=\columnwidth]{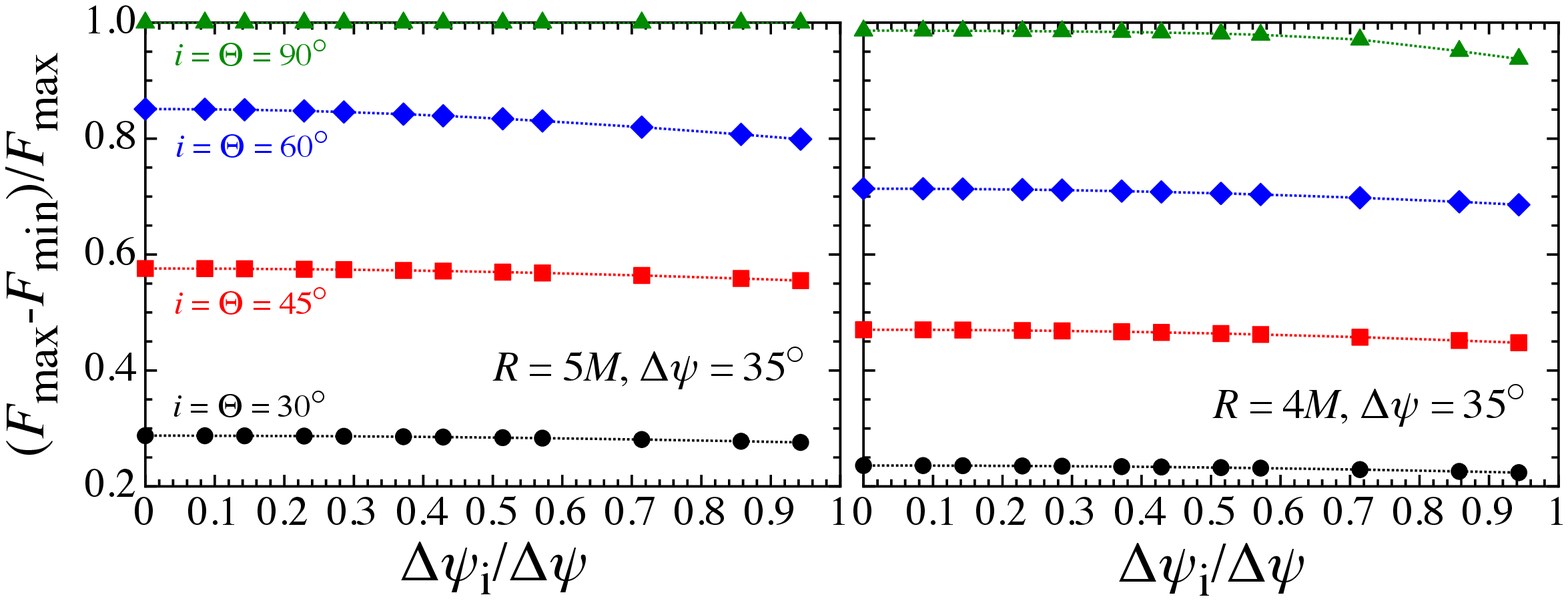}
\end{center}
\caption{
Same as Fig.~\ref{fig:Ratio70}, but for the neutron star model with $\Delta\psi=35^\circ$.
}
\label{fig:Ratio35}
\end{figure}

In Fig.~\ref{fig:RD35}, the light curves from the ring-shaped hot spot with
$\Delta\psi=35^\circ$ are shown for $R=5M$ (upper panel) and for $R=4M$
(lower panel).
In Fig.~\ref{fig:Ratio35}, we show the value of $(F_{\rm max}-F_{\rm
min})/F_{\rm max}$ is as a function of $\Delta\psi_i/\Delta\psi$. Here, we
considered $\Delta\psi_i=0^\circ$, $5^\circ$, $15^\circ$, $25^\circ$, and
$33^\circ$, which correspond the same values of  $\Delta\psi_i/\Delta\psi$
as for $\Delta\psi=70^\circ$ case, i.e., $\Delta\psi_i/\Delta\psi = 0$, $0.14$,
$0.43$, $0.71$, and $0.93$, respectively.
The behavior of the light curves is qualitatively the same as in the case with
$\Delta\psi=70^\circ$, but the $\Delta\psi_i$ dependence is quite weak.
Nevertheless, since the invisible zone becomes larger as the stellar compactness
is smaller, one may have a chance to observe the deviation in the light curve
for the less compact stellar models, although such a deviation must be still very small.
In practice, for $i=\Theta=60^\circ$ one can see a stronger dependence on
$\Delta\psi_i$ in the light curve for the neutron star model with $R=5M$
than that from the model with $R=4M$.
Anyway, if $\Delta\psi$ is less than $35^\circ$, it seems to be difficult to
observationally identify the $\Delta\psi_i$ dependence.  In such a case, via the
observation of the light curve one may be able to discuss the relation between
the stellar compactness and the spot size ($\Delta \psi$) independently of
$\Delta\psi_i$.

\section{Conclusion}
\label{sec:V}

In this work we have calculated bolometric light curves from a single hot spot
on a slowly rotating neutron star, taking into account the effects of the size
and shape of the spot. In order to explore how the light curves are affected by
changes in shape and size, we have specifically considered a filled circle hot
spot and a ring-shaped hot spot.
We found that the light curve from a filled circle hot spot that has an opening
angle of less than $5^\circ$ can be estimated with good accuracy (less than
$1\%$) even with the point-like approximation if the center of the hot spot is
outside the invisible zone.
Moreover, we showed that since the invisible zone at the far side of the star
becomes smaller as the stellar compactness increases, the light curves may not
be eclipsed if the spot size is large enough for large compactnesses. Thus,
through the observation of light curves, one may in principle constrain the
relation between the stellar compactness and the spot size (see
Fig.~\ref{fig:MR-Dcri}).


For annular hot spots, we found that the resulting light curves are
hardly distinguishable from the filled circular case, when the
opening angle of the outer boundary of the hot spot is less than $\sim
35^{\circ}$.
This result yields a conservative typical hot spot size
for which its geometry does not affect the resulting light curve.
This is particular interesting given the fact that hydromagnetic numerical
simulations of accreting flows of matter onto millisecond pulsars
indicate that the resulting hot spot is not necessarily
circular~\cite{Kulkarni:2005cs,Kulkarni:2013sza} with implications
to the modeling of known sources~\cite{Ibragimov:2009js,Kajava:2011fh}
as mentioned previously.

In this study, as a first step, we considered the light curve from a
slowly rotating neutron star in order to isolate the finite size
and geometry influences on the resulting light curves.
To bring our work to the level of astrophysical realism necessary
to analyze, e.g. real NICER data, it would be important to revisit some of our
simplifying assumptions.
For instance, the inclusion of Doppler and aberration effects
due to the star's rotation can affect light curves obtained here
by skewing them towards the left [e.g. in Fig.~\ref{fig:5Mi}] and 
decreasing their amplitudes. On the other hand, delays in the travel time
from photons emitted from different locations within the hotspot are expected
to be small, although it would be of interest to investigate this
for large hot spots where the inclusion of these two effects will enhance the
difference of the resulting light curves relative to those studied here.
It would also be interesting to study the inclusion of a second hot spot
and (in the case of the annular geometry) allow the excised inner circle to be off center with respect to
the outer boundary. These however come at the cost of expanding the parameter
space to be studied.
For rapidly-rotating neutron stars $\omega \gtrsim 1800$ Hz,
the inclusion of the rotation induced quadrupolar deformation of the
star is critical for light curve modelling~\cite{Cadeau:2004gm,Cadeau:2006dc,Morsink:2007tv}.
Finally, one could describe the radiation emitted using a blackbody
spectra.
The inclusion of additional ingredients and an analysis of how they
affect the results presented here will be studied in the future.

\acknowledgments
H.S. acknowledges support in part by Grant-in-Aid for Scientific Research (C)
through Grant No. 17K05458 provided by JSPS.
H.O.S. was supported by NASA grants No. NNX16AB98G, No. 80NSSC17M0041
and thanks Nicol\'as Yunes for discussions.
G.P. acknowledges financial support provided under the European Union's H2020 ERC, Starting Grant agreement no. DarkGRA-757480.
We thank Sharon Morsink for bringing Refs.~\cite{Baubock:2015ixa,Lockhart:2019nch} to our attention.
Part of this work was achieved using the grant of NAOJ Visiting Joint Research
supported by the Research Coordination Committee, National Astronomical
Observatory of Japan (NAOJ), National Institutes of Natural Sciences (NINS).
H.O.S. and G.P. thank the hospitality of the Division of Theoretical Astronomy
of NAOJ.

\bibliographystyle{apsrev4-1}

\end{document}